\documentclass[iop]{emulateapj}
\usepackage{natbib,graphics,graphicx,multirow,pdflscape}
\begin{document}
\title{The Properties of the Progenitor Supernova, Pulsar Wind, and
  Neutron Star inside PWN G54.1+0.3}
\author{Joseph D. Gelfand}
\affil{NYU Abu Dhabi}
\affil{P.O. Box 903, New York, NY, 10276, USA}
\email{jg168@nyu.edu}
\altaffilmark{1}
\altaffiltext{1}{Affiliate Member, Center for Cosmology and Particle
  Physics, New York University, Meyer Hall of Physics, 4 Washington
  Place, New York, NY 10003, USA}
\and
\author{Patrick O. Slane}
\affil{Harvard-Smithsonian Center for Astrophysics}
\affil{60 Garden Street, Cambridge, MA 02138, USA}
\and
\author{Tea Temim}
\affil{Observational Cosmology Lab, Code 665}
\affil{NASA Goddard Space Flight Center}
\affil{Greenbelt, MD 20771, USA}
\altaffilmark{2}
\altaffiltext{2}{CRESST, University of Maryland -- College Park,
  College Park, MD 20742, USA} 

\begin{abstract}
  The evolution of a pulsar wind nebula (PWN) inside a supernova
  remnant (SNR) is sensitive to properties of the central neutron
  star, pulsar wind, progenitor supernova, and interstellar medium.
  These properties are both difficult to measure directly and critical
  for understanding the formation of neutron stars and their
  interaction with the surrounding medium.  In this paper, we
  determine these properties for PWN G54.1+0.3 by fitting its observed
  properties with a model for the dynamical and radiative evolution of
  a PWN inside an SNR.  Our modeling suggests that the progenitor of
  G54.1+0.3 was an isolated $\sim 15 - 20$~M$_\odot$ star which
  exploded inside a massive star cluster, creating a neutron star
  initially spinning with period $P_0 \sim 30 - 80~{\rm ms}$.  We also
  find that $\gtrsim99.9\%$ of the pulsar's rotational energy is
  injected into the PWN as relativistic electrons and
  positrons whose energy spectrum is well characterized by a broken
  power-law.  Lastly, we propose future observations which can both
  test the validity of this model and better determine the properties
  of this source -- in particular, its distance and the initial spin
  period of the central pulsar.
\end{abstract} 

\section{Introduction}
\label{intro}
Stars born with a mass $\ga8~M_\odot$ (e.g., \citealt{heger03}) are
believed to end their lives in a core-collapse supernova powered by
the gravitational collapse of its iron core into a neutron star
(e.g. \citealt{zwicky38}).  In many cases, this collapse creates a
rapidly spinning (initial rotational period $P_0 \ll 1~{\rm s}$)
neutron star with a strong ($B\sim10^{12}$~G) surface magnetic field
observed as a pulsar.  The rotational energy of such neutron stars
powers a magnetized, highly relativistic outflow called a pulsar wind
\citep{goldreich69,arons02}.  The confinements of this outflow creates
a ``termination shock'' (\citealt{kennel84}; see \citealt{gaensler06}
for a recent review), and the post-shock (``downstream'') pulsar wind
creates a pulsar wind nebula (PWN) as it expands into its
surroundings.  When the neutron star is very young, it is located
inside the supernova remnant (SNR) created by the expansion of the
material ejected during the progenitor explosion into the surrounding
interstellar medium (ISM), creating a SNR.  The evolution of this PWN
depends on characteristics of the central neutron star (e.g., its
initial spin period $P_0$), the composition of the post-shock pulsar
wind, and the properties of the progenitor supernova (e.g., the mass
and initial kinetic energy of the supernova ejecta;
\citealt{kennel84b}, \citealt{gelfand09} and references therein) --
quantities difficult to measure directly but vital for understanding
the physics of core-collapse supernovae.

Currently, the best way of measuring the properties of the central
neutron star, its pulsar wind, and progenitor supernova requires
modeling the dynamical and radiative evolution of a PWN inside an SNR
(e.g. \citealt{reynolds84, gelfand09, tanaka10, bucciantini11}, again
see \citealt{gaensler06} for a recent review).  Such models have been
developed, incorporating the effect of the spin-down of the central
neutron star (e.g. \citealt{bucciantini04, gelfand07, gelfand09}), the
evolution of the surrounding SNR as it expands into the ISM
(e.g. \citealt{gelfand07, gelfand09}), and for different properties of
the pulsar winds after being injected into the PWN at the termination
shock (\citealt{volpi08, fang10, bucciantini11}).  In this paper, we
use the model presented in Section \ref{modeldesc} to fit the observed
properties of PWN G54.1+0.3 listed in Section \ref{obsdata}.  The
detection of both an SNR around this PWN \citep{bocchino10,lang10} and
a pulsar PSR J1930+1852 at its center \citep{camilo02,lu07} makes it
especially well-suited for this type of analysis.  We use the Markoff
Chain Monte Carlo (MCMC) routine described in Section \ref{fitdesc} to
explore the possible parameter space and identify degeneracies between
parameters, and in Section \ref{comp} compare our derived properties
of the neutron star, pulsar wind, progenitor supernova, and
surrounding ISM with the results of previous analyses to determine the
impact of our assumptions.  In Section \ref{implications}, we discuss
the implications of these results concerning the progenitor of this
system (Section \ref{progenitor}), the formation of its associated
pulsar (Section \ref{nsform}), and both the production and
acceleration of particles in the pulsar wind (Section \ref{psrwind}).
Finally, in Section \ref{obstests}, we use our model to predict the
results of future observations of this source and discuss their
potential implications.  Lastly, in Section \ref{conclusion}, we
summarize our results.

\section{Evolutionary Model}
\label{modeldesc}
Our model for dynamical and radiative evolution of a PWN inside an SNR
is closely based on that developed by \citet{gelfand09}.  We assume
the rotational luminosity $\dot{E}$ of the central neutron star
evolves as (e.g. \citealt{gaensler06})
\begin{eqnarray}
\dot{E}(t) & = & \dot{E}_0 \left(1+\frac{t}{\tau_{\rm
    sd}}\right)^{-\frac{p+1}{p-1}}, 
\end{eqnarray}
where $t$ is the time since the progenitor supernova, $\dot{E}_0$ is
the neutron star's initial spin-down luminosity, $\tau_{\rm sd}$ is
neutron star's ``spin-down'' timescale, and $p$ is the neutron star's
braking index \citep{gelfand09}, and that all of the rotational energy
of the neutron star is carried away by the pulsar wind generated in
its magnetosphere.  We further assume that, immediately after the
pulsar wind is injected into the PWN at the termination shock, a
constant fraction $\eta_{\rm B}$ of its energy is in the form of
magnetic fields, while the rest $1-\eta_B$ is in the kinetic energy of
electrons and positrons \citep{gelfand09}.  Theoretical studies
predict that, under most physical conditions, the spectrum of these
particles is well described by a relativistic Maxwellian with a
high-energy power-law tail (e.g., \citealt{spitkovsky08, sironi11}).
While this spectrum can reproduce the broadband spectral energy
distribution (SED) of some PWNe (e.g., \citealt{fang10}), we find it
does not work for G54.1+0.3 for constant parameters.  Instead, we use
a broken power-law inject spectrum, which recent simulations (e.g.,
\citealt{sironi11, sironi13}) are able to produce under certain
physical condition, and has been used to reproduce the broadband SED
of this PWN and others in similar work (e.g. \citealt{chevalier05,
  bucciantini11, torres14}).  In this case, the injected particle
spectrum is:
\begin{eqnarray}
\label{eqn:bpl}
n(E) & = & \left\{\begin{array}{ll}
n_{\rm break}\left(\frac{E}{E_{\rm break}}\right)^{-p_1} & E_{\rm min}
< E < E_{\rm break} \\
n_{\rm break}\left(\frac{E}{E_{\rm break}}\right)^{-p_2} & E_{\rm
  break} < E < E_{\rm max} \\
   \end{array} \right. ,
\end{eqnarray}
where $E_{\rm min}$, $E_{\rm break}$, and $E_{\rm max}$ are,
respectively, the minimum, break, and maximum energy of the injected
particles, $p_1$ and $p_2$ are, respectively, the low and high energy
particle indices, $n(E) \Delta E \Delta t$ is the number of electrons
and positrons injected in the PWN between energies $E$ and $E+\Delta
E$ in time $\Delta t$, and $n_{\rm break} \equiv n(E_{\rm break})$.
We calculate $n_{\rm break}(t)$ by requiring that:
\begin{eqnarray}
\label{eqn:norm}
(1-\eta_B)\dot{E} & = & \int\limits_{E_{\rm min}}^{E_{\rm max}} E n(E) dE
\end{eqnarray}
at all times.

To minimize the number of free parameters, we assume that all
parameters related to the properties of the pulsar wind ($\eta_{\rm
  B}$, $E_{\rm min}$, $E_{\rm break}$, $E_{\rm max}$, $p_1$, $p_2$)
remain constant with time -- in contrast to other models which assume
different temporal evolution's for (some) of these parameters.  For
example, \citet{bucciantini11} assume that $E_{\rm max}$ is
proportional to the electric potential of the pulsar's magnetosphere
$\Phi$, while others set $E_{\rm max}$ to the particle energy whose
Larmor radius is the equal to the the radius of the termination shock
(e.g., \citealt{torres14}) or the PWN itself (e.g., \citealt{li10}).
In Section \ref{comp}, we estimate the systematic uncertainty
resulting from these different assumptions by comparing our result to
those derived from different models.  We also note that our model does
not consider the possibility of ions in the pulsar wind, nor the
magnetic reconnection and particle acceleration beyond
(``downstream'') of the termination shock as predicted by recent 3D
simulations of these systems (e.g., \citealt{porth13,porth14}).  These
processes are expected to primarily affect the spectral evolution of
the PWN (e.g., \citealt{olmi14}), and are left for future work.

As done by \citet{gelfand09}, the dynamical evolution of the PWN is
determined by the motion of the surrounding shell of swept-up
material.  This shell is subject to a net force resulting for the
difference in pressure between the PWN and the SNR just outside the
PWN.  We calculate the pressure just outside the PWN using the
procedure described by \citet{gelfand09}, which assumes the initial
density profile of the supernova ejecta is a uniform density inner
core surrounded by an envelope whose density decreases as $\rho
\propto r^{-9}$, where $r$ is the distance from the center of the SNR,
and that the SNR is expanding into a constant density ISM.

As done by \citet{gelfand09}, we calculate the pressure inside the PWN
assuming that both the PWN's magnetic field strength $B_{\rm pwn}$ and
the particle density are spatially uniform -- i.e., using a
``one-zone'' model for the PWN.  We account for both adiabatic and
radiative losses (assumed to be dominated by synchrotron emission and
inverse Compton scattering off Cosmic Microwave Background photons) of
the electrons and positrons inside the PWN.  The spectrum of photons
generated by the radiative losses are calculated using the same
procedure described in \citet{gelfand09}.  To minimize the number of
free parameters in our model, we do not consider emission from
electrons inverse Compton scattering off additional photon fields. We
also do not allow for the escape of particles from the PWN, whose
effect is discussed in recent work (e.g., \citealt{martin12,
  torres14}).  The effects of both assumptions will be discussed in
Section \ref{comp}, when we compare our results to other models which
include one or both of these physical properties.  In total, our model
has twelve free parameters, as listed in Table \ref{tab:bestparfit}.

\section{Observed and Fitted Properties}
\label{properties}
In Section \ref{obsdata}, we present the observed properties of
G54.1+0.3, and in Section \ref{fitdesc} describe the algorithm used to
determining which combinations of model parameters are able to
reproduce them.  Finally, we compare our results with similar work
(Section \ref{comp}).

\begin{table*}[t]
\footnotesize
\caption{The Set of Model Parameters with the Lowest $\chi^2$ ($\chi^2
  \approx 4.10$), Their Predicted Properties of G54.1+0.3, and the
  Observed Values of These Quantities.}
\vspace*{-0.25cm}
\begin{center}
\begin{tabular}{ccccc}
\hline
\hline
\multicolumn{2}{c}{\it Model Input Parameters} & 
\multicolumn{3}{c}{\it Predicted Observables} \\
\hline
Parameter & {\sc Value} & 
Observable & {\sc Observed Value} & {\sc Predicted Value} \\
\hline
$\log(E_{\rm sn}/10^{51}~{\rm ergs})$ & $-0.03$ & $\theta_{\rm snr}$ &
$6\farcm6 \pm 0\farcm4$ & 6.5 \\ 
$\log(M_{\rm ej}/M_\odot)$ & 1.34 & $\theta_{\rm pwn}$ & $1\farcm14
\pm 0\farcm04$ & 1.12 \\ 
$\log(n_{\rm ism}/{\rm cm}^{-3})$ & $-2.29$ & 1.4 GHz Flux Density &
$433\pm30$~Jy & 429~Jy \\ 
Distance (kpc) & 4.90 & 4.7 GHz Flux Density & $327\pm25$~Jy & 329~Jy \\
$p$ & 2.94 & 8.5 GHz Flux Density & $252\pm20$~Jy & 257~Jy \\
$\log(\tau_{\rm sd}/1~{\rm year})$ & 2.90 & $F_{\rm X,2-10}$ &
$(5.43\pm0.035)\times10^{-12}~\frac{\rm ergs}{\rm s~cm^2}$ &
$5.43\times10^{-12}~\frac{\rm ergs}{\rm s~cm^2}$ \\ 
$\log(\eta_{\rm B})$ & $-3.14$ & $\Gamma$ & $2.09\pm0.01$ & $2.09\pm0.002$ \\
$\log(E_{\rm min}/{\rm GeV})$ & 1.05 & 311 GeV Photon Density &
$(1.10\pm0.56)\times10^{-11}~\frac{\rm photons}{\rm cm^2~s~TeV}$ &
$0.80\times10^{-11}~\frac{\rm photons}{\rm cm^2~s~TeV}$ \\ 
$\log(E_{\rm break}/{\rm GeV})$ & 3.45 & 492 GeV Photon Density &
$(4.2\pm1.4)\times10^{-12}~\frac{\rm photons}{\rm cm^2~s~TeV}$ &
$3.1\times10^{-12}~\frac{\rm photons}{\rm cm^2~s~TeV}$ \\ 
$\log(E_{\rm min}/{\rm GeV})$ & 6.98 & 780 GeV Photon Density &
$(1.12\pm0.45)\times10^{-12}~\frac{\rm photons}{\rm cm^2~s~TeV}$ &
$1.21\times10^{-12}~\frac{\rm photons}{\rm cm^2~s~TeV}$ \\ 
$p_1$ & 1.84 & 1.2 TeV Photon Density &
$(6.2\pm1.7)\times10^{-13}~\frac{\rm photons}{\rm cm^2~s~TeV}$ &
$4.9\times10^{-13}~\frac{\rm photons}{\rm cm^2~s~TeV}$ \\ 
$p_2$ & 2.77 & 3 TeV Photon Density &
$(3.9\pm2.1)\times10^{-14}~\frac{\rm photons}{\rm cm^2~s~TeV}$ &
$7.2\times10^{-14}~\frac{\rm photons}{\rm cm^2~s~TeV}$ \\ 
\hline
\hline
\end{tabular}
\end{center}
\label{tab:bestparfit}
\end{table*}

\subsection{Observed Properties}
\label{obsdata}
G54.1+0.3 is one of the best studied PWNe in the Milky Way.
Associated with radio \citep{camilo02} and X-ray \citep{lu07} pulsar
PSR J1930$+$1852, it is also detected across the electromagnetic
spectrum.  This PWN has a similar extent at both radio and X-ray
energies \citep{lu01, lang10}, with a semi-major axis of
$\sim1\farcm25$ and a semi-minor axis of $\sim1\farcm0$
\citep{lang10}.  Since our model assumes a spherically symmetric PWN
(Section \ref{modeldesc}), we set the angular size of the PWN
$\theta_{\rm pwn}$ our model must reproduce to the ``average'' of its
measured semi-minor and semi-major axes, and use these to determine
the 3$\sigma$ lower and upper limits on $\theta_{\rm pwn}$ (Table
\ref{tab:bestparfit}).  We also require our model to reproduce its
volume-integrated radio \citep{lang10}, X-ray \citep{temim10}, and TeV
$\gamma$-ray \citep{acciari10} properties, listed in Table
\ref{tab:bestparfit}.  We do not attempt to reproduce the mid-infrared
(mid-IR) properties of G54.1+0.3 \citep{koo08, temim10} since this
emission is dominated by material shocked and heated by the expanding
PWN.  Because we are using a one-zone model (Section \ref{modeldesc}),
we also do not attempt to any reproduce spatial variations in its
emission (e.g., \citealt{lu01, temim10}).

Lastly, we require our model to reproduce the size of the SNR.  The
SNR around PWN G54.1+0.3 has been detected at both radio
\citep{lang10} and X-ray \citep{bocchino10} energies, each reporting a
somewhat different angular radius $\theta_{\rm snr}$.  To resolve this
discrepancy, we analyzed an archival D-array 1.4 GHz VLA observation
of this PWN, estimating an SNR angular radius of $\approx6\farcm6$.
We then estimated the error on $\theta_{\rm snr}$ by setting 3$\sigma$
upper and lower limits to those reported by \citet{lang10} and
\citet{bocchino10} .

As listed in Table \ref{tab:bestparfit}, our model has to reproduce
twelve different observed quantities -- equal to the number of model
parameters.  As a result, our fit has zero degrees of freedom.  While
the distance $d$ to G54.1+0.3 is a free parameter in our model (Table
\ref{tab:bestparfit}), the fitting algorithm described in
Section \ref{fitdesc} favors $d=4.5-9~{\rm kpc}$, as derived from an
analysis of its H{\sc i} absorption spectrum \citep{leahy08}.

\subsection{Model Fit}
\label{fitdesc}

To derive the physical properties of the central neutron star, pulsar
wind, and progenitor supernova of G54.1+0.3, we use a Metropolis MCMC
algorithm (e.g., \citealt{gelman13}) to determine which combination of
the twelve model parameters $\theta$ described in Section
\ref{modeldesc} best reproduce the twelve observed properties
${\mathcal D}$ discussed in Section \ref{obsdata} and listed in Table
\ref{tab:bestparfit}. To ensure that each trial reproduces the current
spin-down luminosity of $\dot{E}=1.2\times10^{37}~{\rm ergs}$ and a
characteristic age $t_{\rm ch} = 2900~{\rm years}$ PSR J1930$+$1852
inferred from its measured period $P$ and period-derivative $\dot{P}$
\citep{camilo02}, we set the true age $t_{\rm age}$ of G54.1+0.3 to:
\begin{eqnarray}
  t_{\rm age} & = & \frac{2\tau_{\rm ch}}{p-1}-\tau_{\rm sd},
\end{eqnarray}
and the initial spin-down luminosity $\dot{E}_0$ of this pulsar to:
\begin{eqnarray}
  \dot{E}_0 & = & \dot{E} \left(1 + \frac{t_{\rm age}}{\tau_{\rm sd}}
  \right)^{\frac{p+1}{p-1}},
\end{eqnarray}
where $p$ and $\tau_{\rm sd}$ are respectively the pulsar's braking
index and spin-down timescale.

For a given combination, we first determine the model-predicted value
of each observable ${\mathcal M}$.  We then calculate the likelihood
${\mathcal L}({\mathcal D} | \theta)$ this set of parameters
accurately represents the data:
\begin{eqnarray}
  {\mathcal L}({\mathcal D} | \theta) & = & \prod_{i=1}^{12}
  \frac{1}{\sqrt{2\pi}\sigma_i} e^{-\frac{1}{2}
    \left(\frac{\mathcal{M}_i-\mathcal{D}_i}{\sigma_i} \right)^2},
\end{eqnarray}
where $\sigma_i$ is the error on each observed quantity.  The MCMC
algorithm then searches the possible 12-dimensional parameter space
for the combinations with the largest $\ln \mathcal{L}$:
\begin{eqnarray}
  \ln \mathcal{L} & = & \sum_{i=1}^{i=12} \left[  \ln
  \left(\frac{1}{\sqrt{2\pi}\sigma_i} \right) -\frac{1}{2}
  \left(\frac{\mathcal{M}_i-\mathcal{D}_i}{\sigma_i} \right)^2 \right] \\
  & = & -\frac{1}{2}\chi^2 + C,
\end{eqnarray}
where $\chi^2$ is defined as:
\begin{eqnarray}
  \chi^2 & = & \sum_{i=1}^{i=12}
  \left(\frac{\mathcal{M}_i-\mathcal{D}_i}{\sigma_i} \right)^2,
\end{eqnarray}
and $C$ is the same for all combinations.  Therefore, maximizing $\ln
\mathcal{L}$ is equivalent to minimizing $\chi^2$.  It conducts this
search using the following procedure:
\begin{enumerate}
\item For a given combination $\theta_n$, evaluate $\ln
  \mathcal{L}_n$.
\item Propose $\theta_{n+1}$ which is $\theta_n + f(\theta)$, where
  $f(\theta)$ is a set of random, zero-mean, Gaussian distributed
  numbers whose width varies for each model parameter.
\item Calculate $\ln \mathcal{L}_{n+1}$ for the proposed
  $\theta_{n+1}$.
\item If $\frac{\mathcal{L}_{n+1}}{\mathcal{L}_n} \geq \delta$, where
  $\delta$ is a random, Gaussian distributed number between 0 and 1,
  then $\theta_{n+2}$ is calculated with $\theta_{n+1}$ as a starting
  point.  Otherwise, $\theta_{n+2}$ is calculated with $\theta_{n}$ as
  a starting point.
\end{enumerate}
The width of $f(\theta)$ was chosen such that the $\theta_{n+1}$
satisfies the above condition 25\% -- 40\% of the time (Mandel 2015,
private communication; \citealt{gelman13}).

\begin{figure}[t]
  \begin{center}
    \includegraphics[width=0.48\textwidth]{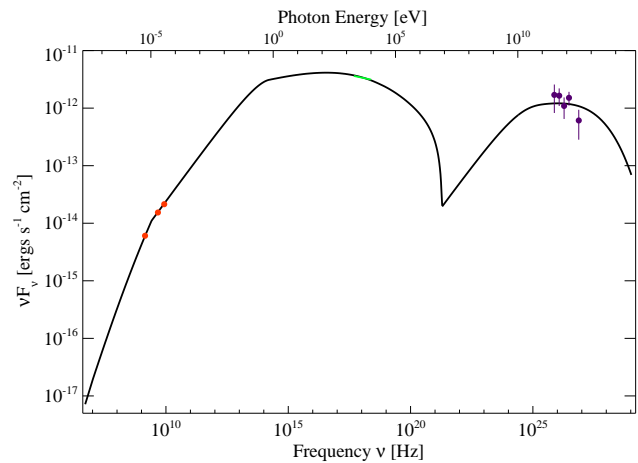} 
  \end{center}
  \caption{The broadband spectral energy diagram of PWN G54.1+0.3
    predicted by the model described in Section \ref{modeldesc} for the
    parameters listed in Table \ref{tab:bestparfit}.  The red, green, and
    purple points are, respectively, the observed radio, X-ray, and TeV
    $\gamma$-ray emission (Section \ref{obsdata}, Table
    \ref{tab:bestparfit}).}
  \label{fig:bestspec}
\end{figure}

To explore a large area of the possible parameter space, we conducted
45 MCMC runs of 50,000 trials, each with different initial parameters.
The initial values were concentrated in regions favored by our current
theoretical ($E_{\rm sn} \sim 10^{51}~{\rm ergs}$ and $M_{\rm ej}
\lesssim 20~{\rm M}_{\odot}$; \citealt{heger03}) and observational ($p
\lesssim 3$; \citealt{livingstone11}) understanding of these sources.
The parameters of the trial with the lowest $\chi^2$ ($\chi^2 \approx
4.10$) are listed in Table \ref{tab:bestparfit}, as are the observed
and predicted properties of G54.1+0.3.  As shown in Figure
\ref{fig:bestspec}, this set of parameters accurately reproduces the
broadband SED of this PWN.

Our search of parameter space allows us to estimate the (statistical)
confidence interval of a given parameter by first ordering, from
lowest to highest, its value in all accepted trials.  The parameter's
90\% confidence interval region is between the 5th and 95th percentile
values in this list (Hogg 2015, private communication;
\citealt{gelman13}).  This range for each parameter is given in Table
\ref{tab:othermodels}, but is sensitive to the chosen distribution of
initial parameters of the MCMC chain.  This bias causes the ``best''
value of $p$ and $\tau_{\rm sd}$ falling outside the quoted ``90\%
confidence interval'' (Table \ref{tab:othermodels}).

\begin{table*}[t]
\caption{Linear Pearson Correlation coefficient $r_{xy}$ (Equation
  (\ref{eqn:correlate})) between the model parameters as calculated for
  all trials with $\chi^2<7.10$. \label{tab:cormatrix}}
\vspace*{-0.25cm}
\scriptsize
\begin{center}
\begin{tabular}{c|cccccccccccc}
\hline
\hline
  & $\log E_{\rm sn}$ & $\log M_{\rm ej}$ & $\log n_{\rm ism}$ & $p$ &
$\log \tau_{\rm sd}$ & $\log \eta_{\rm B}$ & $\log E_{\rm max}$ &
$\log E_{\rm min}$ & $p_1$ & $\log E_{\rm break}$ & $p_2$ & $d$ \\ 
\hline
$\log E_{\rm sn}$ & {\bf 1.00} & {\bf 0.87} & {\bf 0.82} & -0.12 &
0.03 & 0.22 & -0.35 & 0.13 & 0.16 & 0.18 & -0.17 & 0.23 \\ 
$\log M_{\rm ej}$ & {\bf 0.87} & {\bf 1.00} & {\bf 0.59} & -0.18 &
-0.12 & -0.07 & -0.29 & -0.19 & -0.23 & -0.15 & -0.49 & 0.37 \\ 
$\log n_{\rm ism}$ & {\bf 0.82} & {\bf 0.59} & {\bf 1.00} & -0.31 &
0.44 & -0.01 & -0.43 & 0.13 & 0.18 & 0.04 & 0.25 & -0.33 \\ 
$p$ & -0.12 & -0.18 & -0.31 & {\bf 1.00} & {\bf -0.83} & -0.04 & 0.01
& 0.01 & -0.13 & -0.01 & -0.01 & 0.09 \\ 
$\log \tau_{\rm sd}$ & 0.03 & -0.12 & 0.44 & {\bf -0.83} & {\bf 1.00}
& 0.05 & -0.11 & 0.18 & 0.36 & 0.12 & 0.42 & -0.48 \\ 
$\log \eta_{\rm B}$ & 0.22 & -0.07 & -0.01 & -0.04 & 0.05 & {\bf 1.00}
& 0.09 & 0.49 & {\bf 0.68} & {\bf 0.73} & -0.05 & {\bf 0.54} \\ 
$\log E_{\rm max}$ & -0.35 & -0.29 & -0.43 & 0.01 & -0.11 & 0.09 &
{\bf 1.00} & 0.14 & 0.04 & 0.15 & 0.13 & 0.17 \\ 
$\log E_{\rm min}$ & 0.13 & -0.19 & 0.13 & 0.01 & 0.18 & 0.49 & 0.14 &
{\bf 1.00} & {\bf 0.78} & {\bf 0.68} & 0.35 & 0.07 \\ 
$p_1$ & 0.16 & -0.23 & 0.18 & -0.13 & 0.36 & {\bf 0.68} & 0.04 & {\bf
  0.78} & {\bf 1.00} & {\bf 0.93} & 0.40 & 0.12 \\ 
$\log E_{\rm break}$ & 0.18 & -0.15 & 0.04 & -0.01 & 0.12 & {\bf 0.73}
& 0.15 & {\bf 0.68} & {\bf 0.93} & {\bf 1.00} & 0.28 & 0.36 \\ 
$p_2$ & -0.17 & -0.49 & 0.25 & -0.01 & 0.42 & -0.05 & 0.13 & 0.35 &
0.40 & 0.28 & {\bf 1.00} & {\bf -0.71} \\ 
$d$ & 0.23 & 0.37 & -0.33 & 0.09 & -0.48 & {\bf 0.54} & 0.17 & 0.07 &
0.12 & 0.36 & {\bf -0.71} & {\bf 1.00} \\ 
\hline
\hline
\end{tabular}
\end{center}
\vspace*{-0.25cm}
{\it Note}: Values in bold indicates that $|r_{xy} \geq 0.5|$,
indicating a significant degeneracy between the two parameters.
\end{table*}

\begin{figure}[tbh]
  \begin{center}
    \includegraphics[width=0.48\textwidth]{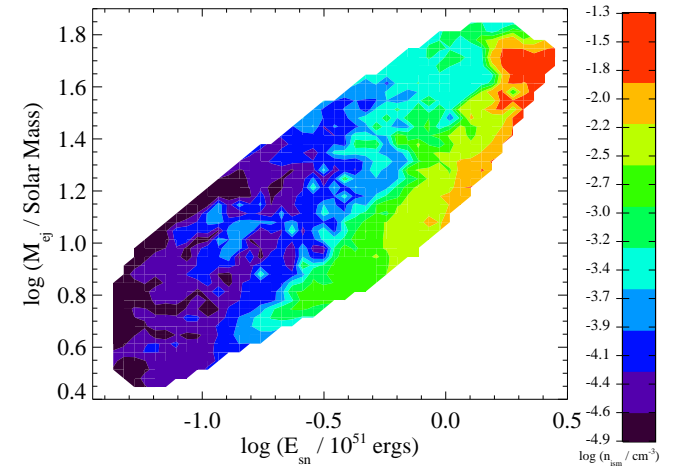} 
  \end{center}
  \caption{The ISM density $n_{\rm ism}$ (color scale) for different
    values of the initial kinetic energy $E_{\rm sn}$ and mass $M_{\rm
      ej}$ of the supernova ejecta for trials with $\chi^2 < 7.10$
    (the 3$\sigma$ parameter space).}
  \label{fig:esn-mej}
\end{figure}

\begin{figure}[tbh]
  \begin{center}
    \includegraphics[width=0.48\textwidth]{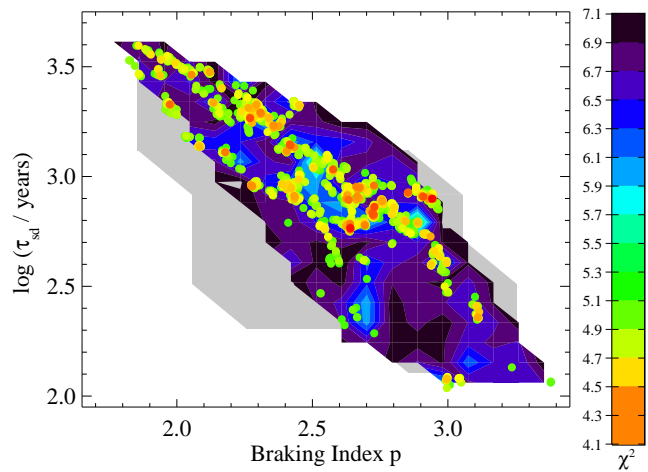}
  \end{center}
  \caption{The $\chi^2$ for trials with different values of the
    braking index $p$ and spin-down timescale $\tau_{\rm sd}$ of PSR
    J1930+1852, with red signifying a lower $\chi^2$ (better fit) and
    black a higher $\chi^2$ (worse fit).  The dots indicate trials
    with $\chi^2 < 5.10$ and are included to better demonstrate the
    degeneracy between these two parameters.  The clumpiness of these
    points primarily reflects the sampling of the parameter space by
    our MCMC algorithm.}
  \label{fig:p-tau}
\end{figure}

\begin{figure}[tbh]
  \begin{center}
    \includegraphics[width=0.48\textwidth]{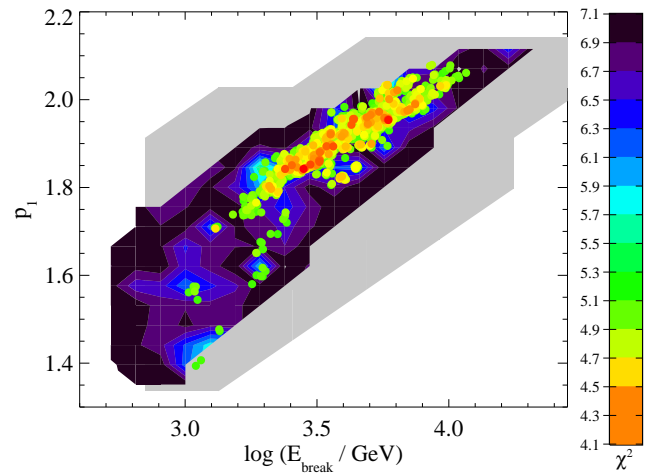} 
  \end{center}
  \caption{The $\chi^2$ for trials with different values of the break
    energy $E_{\rm break}$ and low-energy particle index $p_1$, with
    red signifying a lower $\chi^2$ (better fit) and black a higher
    $\chi^2$ (worse fit).  The dots indicate trials with $\chi^2 <
    5.10$ and are included to better demonstrate the degeneracy
    between these two parameters.  The clumpiness of these points
    primarily reflects the sampling of the parameter space by our MCMC
    algorithm.}
  \label{fig:ebreak-p1}
\end{figure}

\begin{figure}[h]
  \begin{center}
    \includegraphics[width=0.48\textwidth]{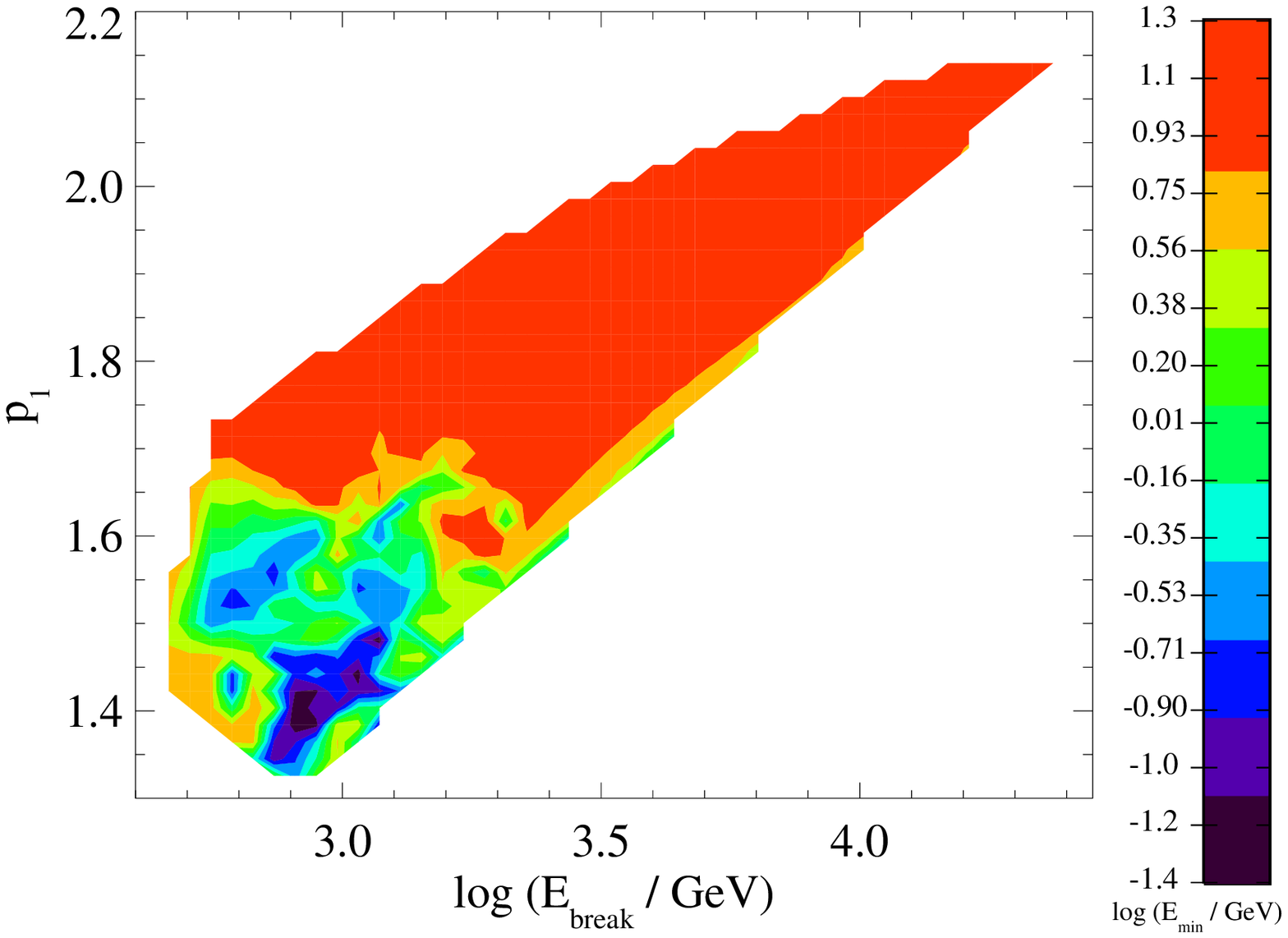} 
  \end{center}
  \caption{The minimum energy $E_{\rm min}$ (color scale) for
    different values of the break energy $E_{\rm break}$ and
    low-energy particle index $p_1$ in the pulsar wind for trials with
    $\chi^2 < 7.10$ (the 3$\sigma$ parameter space).}
  \label{fig:emindist}
\end{figure}

\begin{figure}[h]
  \begin{center}
    \includegraphics[width=0.48\textwidth]{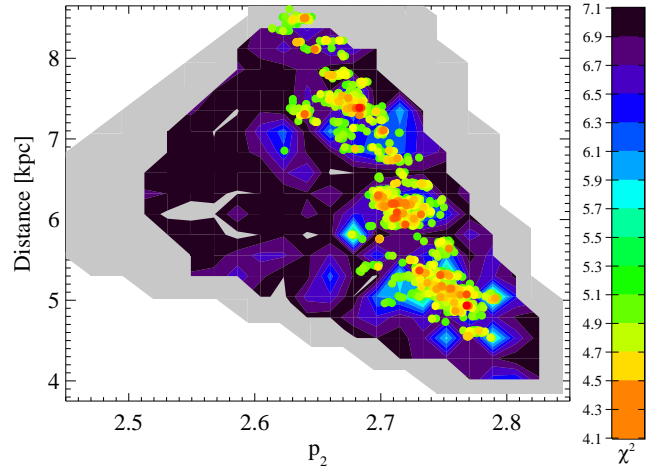} 
  \end{center}
  \caption{The $\chi^2$ for trials with different values of the
    high-energy particle index $p_2$ and distance $d$ to G54.1+0.3,
    with red signifying a lower $\chi^2$ (better fit) and black a
    higher $\chi^2$ (worse fit).  The dots indicate trials with
    $\chi^2 < 5.10$ and are included to better demonstrate the
    degeneracy between these two parameters.  The clumpiness of these
    points primarily reflects the sampling of the parameter space by
    our MCMC algorithm.}
  \label{fig:p2-dist}
\end{figure}

\begin{figure*}[tbh]
  \begin{center}
    \includegraphics[width=0.49\textwidth]{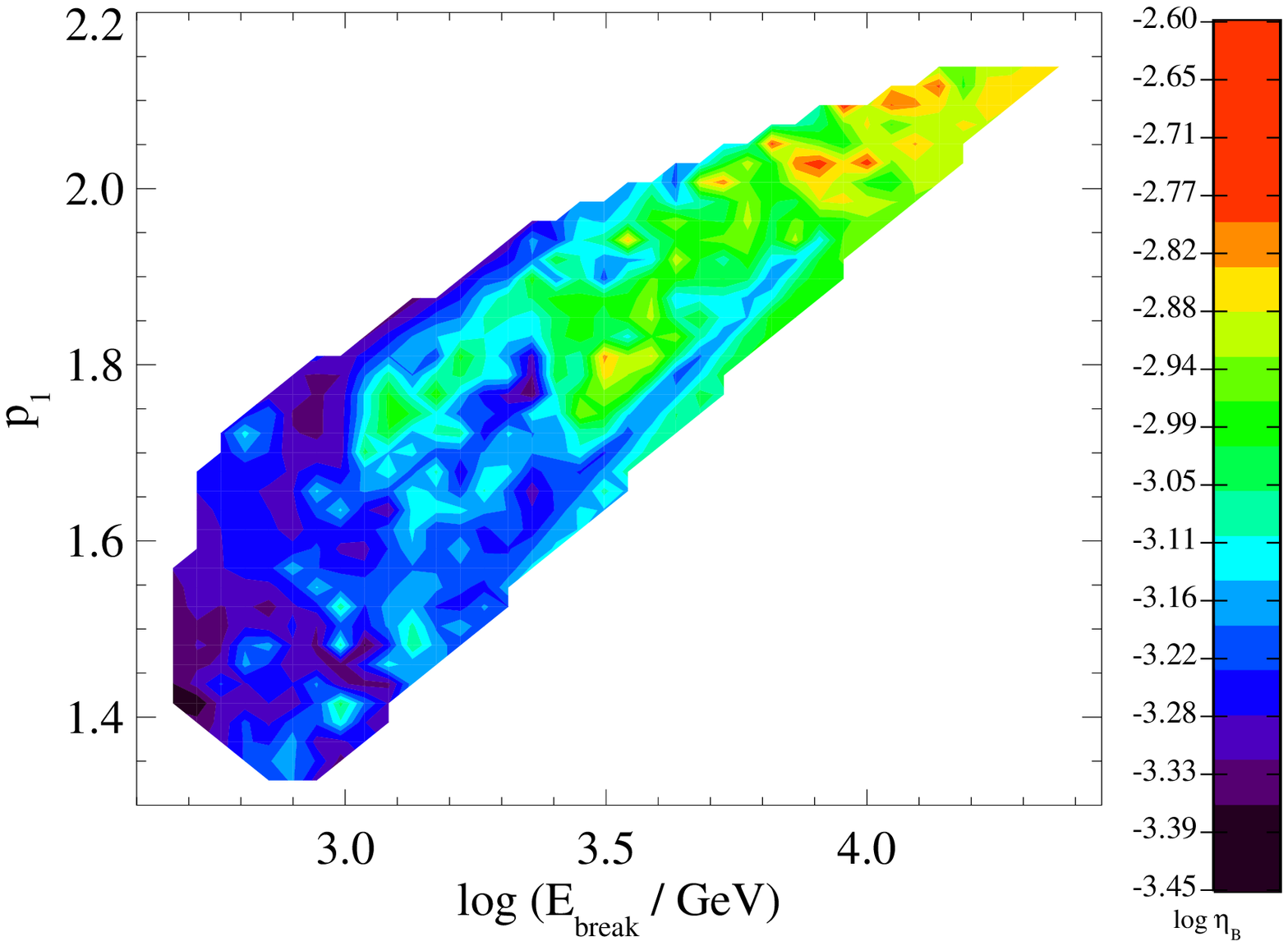} 
    \includegraphics[width=0.49\textwidth]{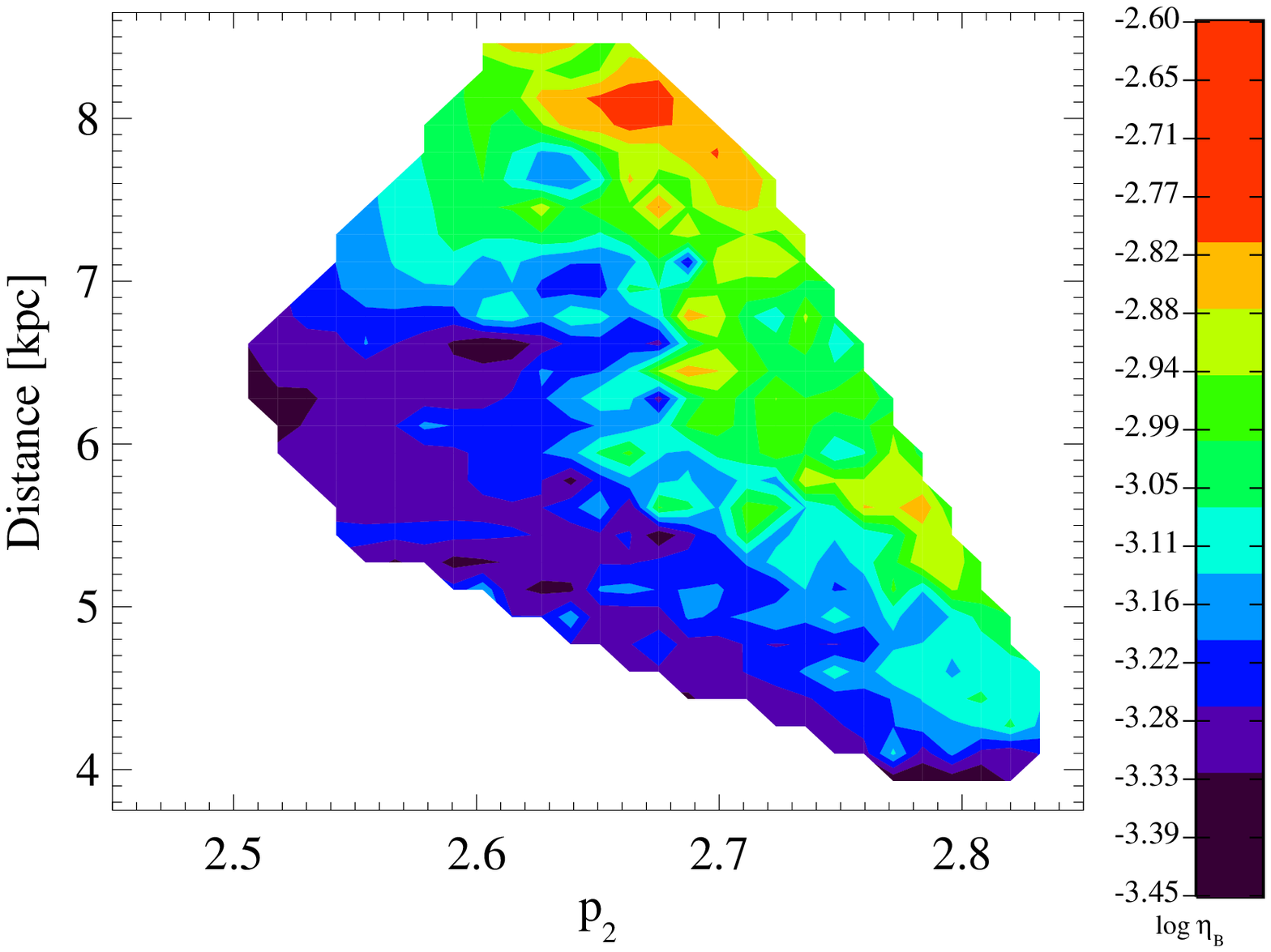} 
  \end{center}
  \caption{The magnetization of the pulsar wind $\eta_{\rm B}$ (color
    scale) for different values of the break energy $E_{\rm break}$
    and low-energy particle index $p_1$ ({\it right}) and distance $d$
    and high energy particle index $p_2$ ({\it left}).  Both are
    calculated for for trials with $\chi^2 < 7.10$ (the 3$\sigma$
    parameter space).}
  \label{fig:dist-etab}
\end{figure*}

Our exploration of the possible parameter space also allows us
identify degeneracies between the various input parameters.  We
calculated the linear Pearson Correlation coefficient $r_{xy}$,
defined to be:
\begin{eqnarray}
  \label{eqn:correlate}
  r_{xy} & = &
  \frac{\sum\limits_{i=1}^{N} (x_i-\bar{x})(y_i-\bar{y})}{\sqrt{ 
      \sum\limits_{i=1}^N(x_i-\bar{x})^2} \sqrt{ 
      \sum\limits_{i=1}^N(y_i-\bar{y})^2 }} ,
\end{eqnarray}
between each pair of model parameters $x$ and $y$, where $\bar{x}$
and $\bar{y}$ are their average values, $x_i$ and $y_i$ are their
values for a particular trial, and $N$ is the number of trials,
using only trials with $\chi^2 < 7.10$ (spanning the 3$\sigma$
parameter space).  If $r_{xy} < 0$, then $x$ and $y$ are inversely
correlated (higher values of $x$ correspond to lower values of $y$),
while if $r_{xy} > 0$, $x$ and $y$ are correlated (higher values of
$x$ correspond to higher values of $y$). Additionally, by
construction, $|r_{xy}| < 1$, with $|r_{xy}| \approx 1$ suggesting
that $x$ and $y$ are strongly correlated while $|r_{xy}| \approx 0$
suggests $x$ and $y$ are weakly correlated.

As shown in Table \ref{tab:cormatrix}, there are significant
degeneracies between various parameters.  For example, the initial
kinetic energy $E_{\rm sn}$ and mass $M_{\rm ej}$ of the supernova
ejecta and the density of the surrounding ISM $n_{\rm ism}$ are
strongly degenerate, with a more energetic supernova explosion
requiring a larger ejecta mass occurring in a denser environment
(Figure \ref{fig:esn-mej}).  A similar degeneracy was reported in a
recent analysis of Kes 75, which discusses possible physical origins
for this behavior \citep{gelfand14}.  The pulsar braking index $p$
and spin-down timescale $\tau_{\rm sd}$ are also strongly
degenerate, with higher values of $p$ requiring lower values of
$\tau_{\rm sd}$ (Figure \ref{fig:p-tau}).  The break energy $E_{\rm
  break}$ in the spectrum of particles injected at the termination
shock strongly depends on the low energy particle index $p_1$, with
higher values of $E_{\rm break}$ requiring a ``softer'' (higher
values of $p_1$) particle spectrum (Figure \ref{fig:ebreak-p1}).
The minimum energy of particles injected at the termination shock
$E_{\rm min}$ also depends on $p_1$ and $E_{\rm break}$ -- $E_{\rm
  min} \approx 10~{\rm GeV}$ for larger values of $p_1$ ($p_1
\gtrsim 1.7$) while lower values of $E_{\rm min}$ require lower
values of $p_1$ (Figure \ref{fig:emindist}).  Additionally, the high
energy particle index $p_2$ is strongly degenerate with the distance
$d$ to G54.1+0.3, with larger distances requiring a ``harder''
(lower values of $p_2$) injection spectrum (Figure
\ref{fig:p2-dist}).  Furthermore, the magnetization of the pulsar
wind $\eta_{\rm B}$ is degenerate with $p_1$, $E_{\rm break}$, and
$d$.  As shown in Figure \ref{fig:dist-etab}, a more magnetized
pulsar wind (higher $\eta_{\rm B}$) requires a higher break energy
$E_{\rm break}$ (and correspondingly higher values of $p_1$) and a
larger distance $d$ (and correspondingly lower values of $p_2$).

\subsubsection{Comparison with Other models}
\label{comp}

In this section, we compare our results with those obtained using
other models for the evolution of a PWN inside a SNR to determine how
our analysis is affected by the assumptions made by our model
described in Section \ref{modeldesc} -- allowing us to estimate the
systematic uncertainty of this approach.  The results of these
different models are provided in Table \ref{tab:othermodels}.

\begin{table*}[tbh]
\caption{The 90\% Confidence Interval of the Properties of G54.1+0.3
  Derived from our Analysis, Compared with Values Derived from
  Previous Analyses of this Source.\label{tab:othermodels}}
\vspace*{-0.25cm}
\begin{center}
\tiny
\begin{tabular}{lcccccc}
\hline
\hline
{\sc Parameter} & This Work & \citet{chevalier05} & \citet{bocchino10}
& \citet{li10} & \citet{tanaka11} & \citet{torres14} \\  
\hline
$E_{\rm sn}$ ($10^{51}$~ergs) & $0.08 - 1.5$ & $\equiv1$ & $0.3-1.6$ &
$\cdots$ & $\cdots$ & $\equiv 1$ \\  
$M_{\rm ej}$ ($M_\odot$) & $5.7-44$ & $\equiv5$ & $\equiv8$ & $\cdots$
& $\cdots$ & $\equiv 20$ \\ 
$n_{\rm ism}$ (cm$^{-3}$) & $(0.03-6.3)\times10^{-3}$ & $\cdots$  & $\sim0.2$ & 
$\cdots$ & $\cdots$ & $\equiv 10$ \\ 
Distance~(kpc) & $4.6-8.1$ & $\sim5$ & $\equiv6.2$ & $\equiv6.2$ &
$\equiv$6.2 & $\equiv 6$ \\
Braking Index $p$ & $1.90-2.93$ & $\equiv3$ & $\equiv3$ & $\equiv3$ &
$\equiv3$ & $\equiv 3$ \\ 
$\tau_{\rm sd}$ (years) & $280-3500$ & $\approx1400$ & $\equiv3$ & $\cdots$ &
600 / 1200 & $1171$ \\ 
$\eta_B$ & $(0.44-2.2)\times10^{-3}$ & $\equiv\frac{3}{7}$ & $\cdots$ &
$\sim1.5\times10^{-3}$ & $0.3\times10^{-3}$ / $2\times10^{-3}$ &
$5\times10^{-3}$ \\   
$E_{\rm min}$ (GeV) & $0.31-15$ & $\cdots$ & $\cdots$ & $\equiv0.05$ &
$<10$ & $\cdots$ \\ 
$E_{\rm break}$ (TeV) & $0.71-11$ & $\cdots$ & $\cdots$ &
$\equiv0.26$ & 0.15 / 0.09 & 0.3 \\ 
$E_{\rm max}$ (PeV) & $0.96-2700$ & $\cdots$ & $\cdots$ & Variable &
$>0.5$ & 0.38 (Variable) \\
$p_1$ & $1.43-2.08$ & $\equiv1.26$ & $\cdots$ & $\equiv1.2$ & $1.2$ & $1.2$ \\
$p_2$ & $2.60-2.78$ & $\equiv2.8$ & $\cdots$ & $\sim2.8$ & $2.55$ & $2.8$
\\
Age [years] & $2100-3600$ & $\approx1500$ & $1800-3300$ & $\sim2900$ &
2300 / 1700 & 1700 \\ 
$\dot{E}_0$ [ergs~s$^{-1}$] & $(0.06 - 2.5)\times10^{39}$ &
$\approx5.1\times10^{37}$ & $\equiv4\times10^{38}$ &
$\equiv1.4\times10^{39}$ & $2.9\times10^{38}$ / $6.9\times10^{37}$ &
$7.2\times10^{37}$ \\  
$P_0$ [ms] & $32-84$ & $\approx100$ & $\equiv56$ & $\cdots$ & 62 / 87
& 87 \\
\hline
\end{tabular}
\end{center}
\vspace*{-0.25cm}
{\it Note}: \citet{chevalier05} do not specify a braking index $p$ for
this neutron star, and the quoted values of $\tau_{\rm sd}$ and
$\dot{E}_0$ are calculated assuming $p\equiv3$ for the age derived in
their analysis.  As described in Section \ref{comp}, \citet{tanaka11}
calculate the properties of this PWN assuming two different energy
densities of the background IR photon field, with the values to the
left of the ``/'' inferred for a lower energy density while the values
to the right are those inferred for a higher energy density.
\end{table*}

\citet{chevalier05} uses the measured spectral properties and radius
of this PWN and the spin-down properties of the central pulsar to
primarily estimate the birth properties of the neutron star, assuming
$M_{\rm ej} \equiv 5~M_\odot$ and $E_{\rm sn}=10^{51}~{\rm ergs}$ -- a
combination not favored by our fits (Figure \ref{fig:esn-mej}).  He
did not attempt to reproduce the broadband SED, and set $p_1$ and
$p_2$ to values inferred from single power-law fits to the observed
radio and X-ray spectrum.  While the value of $p_2$ derived from this
method agrees with our value, the value of $p_1$ does not since, {\bf
  for $E_{\rm min} \approx 10~{\rm GeV}$}, the SED predicted by our
model contains a spectral break between 1.4 and 4.8 GHz (Figure
\ref{fig:bestspec}).  Additionally, he assumes that $\eta_{\rm B} =
\frac{3}{7}$ \citep{chevalier05}, significantly higher value than
allowed by our fits.  The higher value of $\eta_{\rm B}$ decreases the
particle energy inside the PWN, resulting in an initial period $P_0$
significantly higher than we derive.

\citet{bocchino10} infer the age and the properties of both the
progenitor supernova and surrounding ISM from the X-ray emission
associated with the SNR shell.  They derive $n_{\rm ism} \sim
0.2$\ cm$^{-3}$ assuming a distance of $d\equiv 6.2~{\rm kpc}$, higher
than the values preferred by our modeling (Table
\ref{tab:othermodels}).  Their derived age $t_{\rm age}$ and supernova
explosion energy $E_{\rm sn}$ are sensitive to the ratio of the
electron and ion temperature in the SNR, with $E_{\rm sn}=(0.3-0.7)
\times 10^{51}~{\rm ergs}$ and $t_{\rm age} \sim 2500-3300~{\rm
  years}$ if the electrons and ions are in equipartition, while
$E_{\rm sn} = (0.5-1.6) \times 10^{51}~{\rm ergs}$ and $t_{\rm age}
\sim 1800-2400~{\rm years}$ if the ions are $\sim2\times$ hotter than
the electrons.  Both sets of $E_{\rm sn}$ and $t_{\rm age}$ are
consistent with our results (Table \ref{tab:othermodels}).  They also
found that $M_{\rm ej}=8~M_{\odot}$, $p=3$, and $\tau_{\rm
  sd}=500~{\rm years}$ can reproduce the radius of the PWN and SNR
\citep{bocchino10} -- in agreement with our results.  Since they did
not attempt to reproduce the broadband SED of this source, this
analysis does not constrain the magnetization or spectrum of particles
injected into the PWN at the termination shock.

G54.1+0.3 was also analyzed by \citet{tanaka11}, who reproduce both
the size and broadband SED of this PWN using a model very similar to
ours (Section \ref{modeldesc}) but include inverse Compton scattering of
electrons off photon fields other than the CMB: an optical
($T=4000$~K) photon field with an energy density $u_{\rm opt}=0.5~{\rm
  eV~cm}^{-3}$, and an IR ($T=40$~K) photon field with an energy
density $u_{\rm ir}=0.5~{\rm eV~cm}^{-3}$ or $u_{\rm ir}=2.0~{\rm
  eV~cm}^{-3}$ -- finding that $\eta_{\rm B}$, $E_{\rm break}$, and
the parameters associated with the energetics of the neutron star
($\tau_{\rm sd}$, $t_{\rm age}$, $\dot{E}_0$, and $P_0$) depend on
$u_{\rm ir}$ \citep{tanaka11}.  As listed in Table
\ref{tab:othermodels}, in general our parameters agree -- though their
analysis favors a higher value of $P_0$ (less energetic neutron star)
due to the inclusion of these additional photon fields.

Similar results were obtained by \citet{torres14}, which uses an
evolutionary model that includes the diffusion of particles both
inside and out of the PWN \citep{martin12}.  Like \citet{tanaka11},
they include emission from electrons inverse Compton scattering off
two photon fields in addition to the CMB, one with $T_{\rm
  FIR}=20~{\rm K}$ and energy density $u_{\rm FIR}=2.0~{\rm
  eV~cm}^{-3}$ and the other with $T_{\rm NIR} = 3000~{\rm K}$ and
energy density $u_{\rm FIR}=1.1~{\rm eV~cm}^{-3}$ \citep{torres14} --
again deriving a lower $\dot{E}_0$ (higher $P_0$) than our analysis.
This model also assumes the maximum energy of particles is limited by
confinement in the termination shock -- finding that the current value
of $E_{\rm max}$ is similar to what we require for our model.

Lastly, we compare our results with those of \citet{li10}, who model
the broadband SED of G54.1+0.3 for both a leptonic and combined
leptonic and hadronic origin for the observed $\gamma$-rays.  Like
\citet{torres14}, they allowed the maximum energy of particles
injected at the termination shock $E_{\rm max}$ to vary, setting it to
the energy whose Larmor radius is the radius of the PWN \citep{li10}.
Their model also allows leptons to escape from the PWN, and that these
particles inverse Compton scatter off the CMB, background IR and
optical photons from the Milky Way, and emission from the IR ``loop''
and its embedded point sources around this PWN \citep{koo08, temim10}.
In the purely leptonic case, \citet{li10} derive similar values of
$\eta_{\rm B}$ and $p_2$ despite assuming very different values of
$E_{\rm min}$, $E_{\rm break}$, and $p_1$ (Table
\ref{tab:othermodels}).  

\section{Fit Implications}
\label{implications}
As described in Section \ref{intro}, the derived properties of the
supernova ejecta, surrounding ISM, pulsar, and pulsar wind presented
in Section \ref{fitdesc} allow us to estimate the properties of the
stellar progenitor (Section \ref{progenitor}) the birth properties of
the central pulsar (Section \ref{nsform}), and provide insight to the
generation and acceleration of particles in the pulsar wind (Section
\ref{psrwind}).

\subsection{Progenitor Star}
\label{progenitor}

The initial kinetic energy $E_{\rm sn}$ and mass $M_{\rm ej}$ ejected
in a core-collapse supernova depends on the initial mass, metallicity,
and evolution of the progenitor star (e.g., \citealt{heger03}).
G54.1+0.3 has a galactocentric radius ($\sim6.5-7.5$~kpc for the
favored distance of $d\sim5-8$~kpc) similar to the Sun's
($\sim8-8.5$~kpc; \citealt{andrievsky02b,andrievsky02}), suggesting
its progenitor had approximately Solar metallicity.  A massive star in
a binary is expected to transfer much of their mass to their companion
before it explodes, resulting in a very low ejecta mass (e.g., $M_{\rm
  ej}\lesssim3~M_\odot$; e.g., \citealt{woosley02}).  Since our model
suggests that $M_{\rm ej} \gtrsim 3~{\rm M}_\odot$ for even low energy
explosions (Figure \ref{fig:esn-mej}), we assume the progenitor was
isolated.  Such stars produce a neutron star when they (e.g.,
\citealt{woosley02}):
\begin{enumerate}
\item have an initial mass of $\sim8-20~M_\odot$, in which case they
  explode as a red super-giant, ejecting a lot of material
  ($\gtrsim6-15~M_\odot$), or
\item have extremely high ($\sim50~M_\odot$) initial mass but explodes
  after shedding much of this mass as a Wolf-Rayet star, resulting in
  a low ($\lesssim3~M_\odot$) ejecta mass.
\end{enumerate}
As shown in Figure \ref{fig:esn-mej}, a ``canonical'' supernova
explosion energy of $E_{\rm sn} \sim (0.3-1) \times 10^{51}~{\rm
  ergs}$ requires a higher ejecta mass ($M_{\rm ej} \ga 10~M_\odot$;
Figure \ref{fig:esn-mej}).

\begin{figure}[tb]
  \begin{center}
    \includegraphics[width=0.48\textwidth]{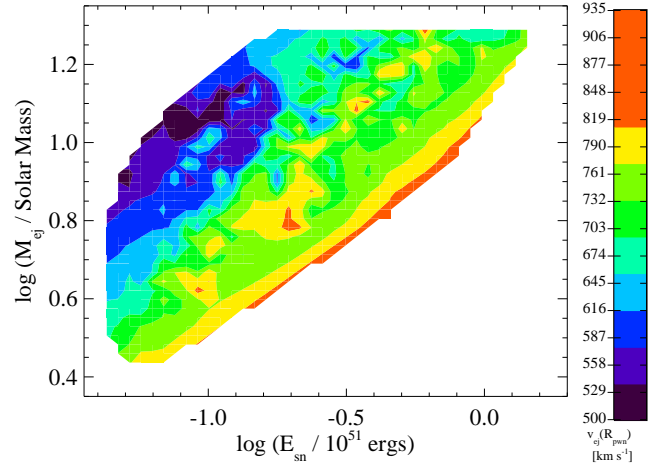} 
  \end{center}
  \caption{The expected expansion speed of the ejecta just outside the
    PWN $v_{\rm ej}(R_{\rm pwn})$ for different values of the
    initial kinetic energy $E_{\rm sn}$ and mass $M_{\rm ej}$ of the
    supernova ejecta for trials with $\chi^2<7.10$, $M_{\rm ej} <
    20~M_{\odot}$, and $p<3$. }
  \label{fig:esn-mej_mswpwn}
\end{figure}

We can further constrain these parameters using the properties
inferred from an analysis of the IR spectrum of the material
surrounding the PWN \citep{koo08, temim10}.  This material is
primarily supernova ejecta, suggesting the SNR ejecta has not yet
mixed with the swept-up and shocked ISM -- consistent with the lack of
collision between the PWN and SNR reverse shock, as required by our
model.  The observed width of the IR lines suggests the surrounding
ejecta are expanding with a speed $v_{\rm ej}(R_{\rm
  pwn})\lesssim500~{\rm km~s}^{-1}$, consistent with $M_{\rm ej}
\sim10-15~M_{\odot}$ of material ejected in a somewhat under-energetic
$E_{\rm sn} \sim (0.1-0.2)\times10^{51}~{\rm erg}$ explosion
\ref{fig:esn-mej_mswpwn}).  Stellar evolution models suggest a
$\sim15-20~M_\odot$ progenitor is required to produce this much ejecta
(e.g., \citealt{heger03}).  This progenitor mass is further supported
by the identification of O and B stars embedded inside the SN ejecta
dust surrounding this PWN \citep{temim10}.  Therefore, G54.1+0.3 was
likely produced by the core-collapse of a $\sim15-20~M_\odot$ star in
a massive star cluster -- possibly the most massive member of this
cluster, and therefore the first to explode.

This progenitor mass, and association with an massive star cluster,
can explain the low ISM density $n_{\rm ism}$ required by our model
(Table \ref{tab:othermodels}).  The winds of main-sequence massive
stars are thought to create low-density bubble with a radius $R_{\rm
  b}$ \citep{chen13}:
\begin{eqnarray}
  R_{\rm b} & = & \left[ (1.22\pm0.04) \frac{M}{M_\odot} - (9.16\pm1.77)\right]
  \left( \frac{P_{\rm ism}/k_{\rm B}}{10^5~{\rm cm^{-3}~K}}
  \right)^{-\frac{1}{3}}~{\rm pc} \\
  & \sim & 7 - 18~\left( \frac{P_{\rm ism}/k_{\rm B}}{10^5~{\rm cm^{-3}~K}}
  \right)^{-\frac{1}{3}}~{\rm pc},
\end{eqnarray}
where $P_{\rm ism}$ is the pressure of the medium outside the wind
bubble and $k_{\rm B}$ is Boltzmann's constant.  For a distance of
$\sim4.5-9$~kpc \citep{leahy08}, this bubble will have an angular size
of $\theta_{\rm b} \sim 2\farcm7 - 14^\prime$.  Winds from the
additional massive stars in the cluster will only increase the size of
this bubble, increasing the likelihood that the SNR is expanding
inside a low density environment.

\subsection{Neutron Star Formation and Evolution}
\label{nsform}

The birth properties of a neutron star reflect the physics of its
formation.  The initial spin period $P_0$ and surface magnetic field
of the neutron star depend on the properties of its progenitor,
particularly the rotation rate of its iron core (e.g.,
\citealt{ott06}), and instabilities active during the supernova
explosion (e.g., \citealt{blondin07, endeve10}), while the spin-down
properties of the neutron star (e.g., its braking index $p$ and
spin-down timescale $\tau_{\rm sd}$) likely depends on its internal
structure (e.g., \citealt{ho12}).  While the theory connecting these
parameters to the underlying physics is far from settled, measuring
these quantities provides important information on these processes.
For example, if the initial rotation of the neutron star is limited by
gravitational waves resulting from r-mode instabilities generated by
the ``fallback'' of material during the supernova onto the
proto-neutron star, the $B_{\rm ns} = 1.0\times10^{13}~{\rm G}$ dipole
surface magnetic field strength inferred from the timing properties of
PSR J1930$+$1852 \citep{camilo02} requires $P_0 \sim 30 - 80~{\rm ms}$
\citep{watts02} -- consistent with the range favored by our model
(Table \ref{tab:othermodels}).

\subsection{Pulsar Wind}
\label{psrwind}

The rotation of the neutron star generates a strong electric potential
(voltage) $\Phi$ at its magnetic poles responsible for both creating
particles in its magnetosphere (e.g., \citealt{goldreich69}).  The
pulsar wind consists of particles which exit the magnetosphere along
open field lines, expected to occur at a minimum rate $\dot{N}_{\rm
  GJ}$
\begin{eqnarray}
  \dot{N}_{\rm GJ} & = & \frac{c\Phi}{e} = 7.6\times10^{33} \left(
  \frac{I_{45}}{P_{33}^3} \frac{\dot{P}}{4\times10^{-13}~{\rm s/s}}
  \right)^{\frac{1}{2}}~{\rm s^{-1}},
\end{eqnarray}
\citep{goldreich69, bucciantini11} where the neutron star's moment of
inertia is $I = I_{45}\times10^{45}~{\rm g~cm^2}$, $P_{33} = P/33~{\rm
  ms}$, and $\dot{P}$ is the neutron star's period-derivative.
However, how particles are both created and leave the neutron star
magnetosphere is poorly understood.

If particles are neither created nor destroyed between the light
cylinder and the termination shock, we can calculate the rate
particles leave the magnetosphere $\dot{N}$ for a particular trial
using Equations \ref{eqn:bpl} and \ref{eqn:norm}.  Our assumption that
the parameters regulating the spectrum of particles injected at the
termination shock ($E_{\rm min}$, $E_{\rm break}$, $E_{\rm max}$,
$p_1$, and $p_2$; Table \ref{tab:bestparfit}) are constant results in
$\dot{N} \propto \dot{E}$ over the life time of the PWN.  As result,
in our model the multiplicity of the pulsar wind $\kappa$
\begin{eqnarray}
  \label{eqn:kappa}
  \kappa & \equiv & \frac{\dot{N}}{\dot{N}_{\rm GJ}},
\end{eqnarray}
varies with time.  Therefore, in addition to calculating the current
multiplicity $\kappa_{\rm now}$, we also calculate the time-integrated
multiplicity $\kappa_{\rm int}$ (e.g., \citealt{dejager07}):
\begin{eqnarray}
  \label{eqn:kappaint}
  \kappa_{\rm int} & = & \frac{\int\limits_0^{t_{\rm age}} \dot{N}
    dt}{\int\limits_0^{t_{\rm age}} \dot{N}_{\rm GJ} dt}.
\end{eqnarray}
Our analysis of G54.1+0.3 indicates that $\kappa_{\rm now} \approx
10^3$ $\kappa_{\rm int} \sim (1-3)\times10^5$ -- both in good
agreement with the values derived from similar analyses of other PWNe
(e.g., \citealt{dejager07}), but higher than that predicted by current
theoretical models (e.g., \citealt{hibschman01}).  Since our model
requires that $p_1 > 0$ and $p_2 > 0$ (Table \ref{tab:othermodels}),
the estimated multiplicity strongly depends on the minimum particle
energy $E_{\rm min}$ injected in the PWN at the termination shock.
Our model suggests $E_{\rm min} \approx 10~{\rm GeV}$ by producing a
``break'' in the radio spectrum around 4.8~GHz (Section \ref{obsdata}; Table
\ref{tab:bestparfit}, Figure \ref{fig:bestspec}).  In
Section \ref{obstests}, we suggest observations which will determine if this
minimum energy and multiplicity are an artifact of having zero degrees
of freedom.

Near the neutron star, the pulsar wind is expected to be highly
magnetized ($\eta_{\rm B} \approx 1$).  However, our model requires
that $\eta_{\rm B} \sim 10^{-3}$ (Table \ref{tab:othermodels}) when
the pulsar wind is injected into the PWN -- requiring that magnetic
energy is converted to particle energy between the neutron star's
light cylinder and the termination shock (e.g., \citealt{kirk03}).
Currently, magnetic reconnection in this region is thought to
transform the pulsar wind from a strongly magnetized to a weakly
magnetized outflow (e.g., \citealt{kirk03, sironi11, sironi14}).
Efficient magnetic reconnection requires that \citep{kirk03}:
\begin{eqnarray}
  \label{eqn:mucrit}
  \mu & < & 3 \left(\frac{\pi^3 e^2}{m_e^2c^5} \dot{E} \right)^{\frac{1}{4}}
\end{eqnarray}
where $e$ and $m_e$ are, respectively, the charge and mass of a
positron, $c$ is the speed of light, and $\mu$, the energy per unit
mass energy of the pulsar wind, is \citep{kirk03}:
\begin{eqnarray}
  \label{eqn:mu}
  \mu & \equiv & \frac{\dot{E}}{\dot{N}mc^2},
\end{eqnarray}
equivalent to the bulk Lorentz factor of the pulsar wind $\gamma_{\rm
  w}$ before it reaches the termination shock (i.e., ``upstream'' from
the shock).  Therefore, magnetic reconnection is viable as long
as the spin-down luminosity of PSR J1930$+$1852 is:
\begin{eqnarray}
  \dot{E} & > & \frac{m_e^2 c^5 \mu^4}{81\pi^3 e^2} \approx
  (0.2-1.3)\times10^{35}~\frac{\rm ergs}{\rm s},
\end{eqnarray}
for $\mu \approx (1.5-2.5)\times10^5$ as favored by our model.  Since
this critical $\dot{E}$ is well below its current $\dot{E} \approx
1.7\times10^{37}~\frac{\rm ergs}{\rm s}$ \citep{camilo02}, magnetic
reconnection should occur in the pulsar wind before it reaches the
termination shock -- possibly explaining the weakly magnetized pulsar
wind required by our model.

Recent numerical simulations suggest that magnetic reconnection is the
pulsar wind will produce particles whose spectrum is well described by
a power-law with particle index $p \lesssim 2$, as required by our
model for $E < E_{\rm break}$ (Table \ref{tab:othermodels}), up to an
energy :
\begin{eqnarray}
  \label{eqn:emaxrecon}
  E_{\rm max,recon} & \sim & m_ec^2
  \left[\frac{(\sigma_{\rm recon}+1)(2-p)}{(p-1)}\right]^{\frac{1}{2-p}},
\end{eqnarray}
if the ratio of magnetic to particle energy in the magnetic
reconnection region is $\sigma_{\rm recon} \gtrsim 10$
\citep{sironi14}.  We can test if this is plausible calculating
$\sigma_{\rm recon}$ if, in Equation \ref{eqn:emaxrecon}, $p=p_1$ and
$E_{\rm max,recon} = E_{\rm break}$:
\begin{eqnarray}
  \label{eqn:sigmarecon}
  \sigma_{\rm recon} & \sim & \frac{p_1 - 1}{2-p_1} \left(\frac{E_{\rm
      break}}{m_ec^2}\right)^{2-p_1} - 1.
\end{eqnarray}
For the preferred values of $E_{\rm break}$ and $p_1$ (Table
\ref{tab:othermodels}, Figure \ref{fig:ebreak-p1}), we find that
 $\sigma_{\rm recon} \sim 30 - 10^5$ -- suggesting that magnetic
reconnection in the pulsar wind between the light cylinder and
termination shock could explain the low energy component of the
injected particle spectrum.

We can also use our results to test models for the origin of the high
energy component in the injected particle spectrum.  One possibility
is that $E_{\rm max} = e\Phi$ (e.g., \citealt{bucciantini11}), where
$\Phi$ is the voltage of the pulsar's magnetosphere:
\begin{eqnarray}
  \label{eqn:voltage}
  \Phi & = & \sqrt{\frac{\dot{E}}{c}}.
\end{eqnarray}
The current spin-down luminosity $\dot{E}$ of PSR J1930$+$1852
\citep{camilo02} would suggest that $E_{\rm max} \approx 6~{\rm PeV}$
in its magnetosphere -- consistent with the values $E_{\rm max}$
required by our modeling (Table \ref{tab:othermodels}).  Another
possibility is that these particles are created by additional
acceleration at the termination shock.  Simulations suggest that
efficient acceleration of an electron-positron plasma in this region
requires $\eta_{\rm B} \lesssim 10^{-3}$ (e.g., \citealt{sironi13}),
again consistent with the range of values favored by our modeling.
The maximum particle energy is expected be limited by either
synchrotron cooling or diffusion away from the termination shock, with
the theoretic maximum energy $E_{\rm max,theory}$ being the lower of
the two.  For the pulsar wind properties favored by our modeling, the
maximum energy of the particles accelerated at the termination shock
is limited by diffusion, such that:
\citep{sironi13}:
\begin{eqnarray}
  \label{eqn:maxconf}
  E_{\rm max,theory} & \simeq & 1.9\times10^7 m_ec^2
  \left(\frac{\dot{E}}{10^{38.5}~{\rm \frac{ergs}{s}}}\right)^{\frac{3}{4}}
  \left(\frac{\dot{N}}{10^{40}~{\rm s}^{-1}} \right)^{-\frac{1}{2}} \\
  & \sim & 15-25~{\rm PeV}.
\end{eqnarray}
Since $\sim50\%$ of our trials have $E_{\rm max} < E_{\rm
  max,theory}$, our results are also consistent with highest energy
particles being produced at the termination shock.

Lastly, numerical simulations suggest the spectral shape of particles
injected into the PWN at the termination shock depends strongly on the
structure of the unshocked pulsar wind (e.g., \citealt{sironi11}).
When it leaves the neutron star magnetosphere, the pulsar wind is
expected to be primarily equatorial and composed of regions of
alternating magnetic field directions (e.g., \citealt{bogovalov99}) of
width $\lambda$.  The shape of the resultant particle spectrum is
expected to depend on (\citealt{sironi11}):
\begin{eqnarray}
  \label{eqn:recon}
  \frac{\lambda}{r_L \sigma} & \simeq & 4\pi\kappa \frac{R_{\rm
      LC}}{R_{\rm TS}},
\end{eqnarray}
where $r_L$ and $\sigma$ are, respectively, the relativistic Larmor
radius and magnetization of the unshocked pulsar wind, $\kappa$ is the
multiplicity (Equation \ref{eqn:kappa}), $R_{\rm TS}$ is the radius of
the termination shock, $R_{\rm TS}$ is the radius of termination
shock, and $R_{\rm LC}$ is the radius of the light cylinder:
\begin{eqnarray}
  R_{\rm LC} & = & \frac{cP}{2\pi}.
\end{eqnarray}
Specifically, $\lambda/(r_L \sigma) \gtrsim 10$ is required for the
spectrum of particles accelerated at the termination shock to resemble
the broken power-law required by our model, otherwise it should be
well approximated by a relativistic Maxwellian incompatible with our
analysis (Section \ref{modeldesc}).

We can test this prediction using our trial parameters and the
observed properties of this system.  The measured $P\approx136.86~{\rm
  ms}$ and $\dot{P} \approx 7.51\times10^{-13}~{\rm s/s}$ of PSR
J1830+1852 \citep{camilo02} suggests that currently $\dot{N}_{\rm GJ}
\approx 4.69\times10^{34}~{\rm s}^{-1}$ and $R_{\rm LC} \approx
6.53\times10^8~{\rm cm}$.  Additionally, analysis of a {\it Chandra}
observation identified a ring with semi-major axis $\theta_{\rm TS} =
5\farcs7$ centered on the pulsar, which is believed to mark the
position of the termination shock in this PWN \citep{lu02, temim10}.
For these values, the trial parameters with the lowest $\chi^2$ favor
$\frac{\lambda}{r_L \sigma} \sim 10^{-5} - 10^{-4}$, in contradiction
with the results of \citet{sironi11}.

\section{Observational Tests}
\label{obstests}

While our evolutionary model for a PWN inside an SNR (Section
\ref{modeldesc}) reproduces the observed properties of G54.1+0.3 for a
wide range in parameter space (Table \ref{tab:othermodels}), it is
important to test the validity of this model by predicting the value
of additional observable properties.  Thanks to our parameter
exploration, not only can we predict the values of future
observations, we can also estimate the resulting improvement in the
allowed physical parameters.  For these predictions, we only use
trials with $\chi^2<7.10$, $M_{\rm ej} < 20~M_\odot$, and $p<3.0$.  We
only consider trials with $M_{\rm ej} < 20 M_{\odot}$ since stellar
evolution models suggests this is the maximum ejecta mass possible for
a Solar metallicity star (\citealt{woosley02, heger03}; Heger 2015,
private communication), and only trials with $p<3$ since $p>3$ has yet
to be measured from any isolated neutron star (e.g.,
\citealt{livingstone11}).

Our model can predict properties of the SNR around G54.1+0.3 not yet
measured, for example its expansion velocity $v_{\rm snr}$.  Due to
the young age and low ISM density preferred by our model, we predict
an extremely fast $v_{\rm snr} \sim 3000~{\rm km~s}^{-1}$ -- among the
highest measured or inferred for any other SNR (e.g.,
\citealt{ghavamian07}).  This suggests the identified radio and X-ray
shell may not actually be a SNR but the progenitor's stellar wind
bubble (Section \ref{progenitor}).  This can be determined by the
measuring its radio spectral index ($\alpha$, where flux density
$S_\nu \propto \nu^\alpha$), since the free-free emission expected to
dominate the radio emission from a stellar wind bubble has $\alpha
\gtrsim 0$ while SNRs typically have $\alpha \sim -0.7$.  If future
studies indicate this is a stellar wind bubble, our model would still
favor a $\sim15-20~{\rm M}_\odot$ progenitor based on the properties
of the IR emission around the PWN (Section \ref{progenitor}), but
would offer much weaker constraints on the density of the surrounding
ISM.

\begin{figure}[tb]
  \begin{center}
    \includegraphics[width=0.48\textwidth]{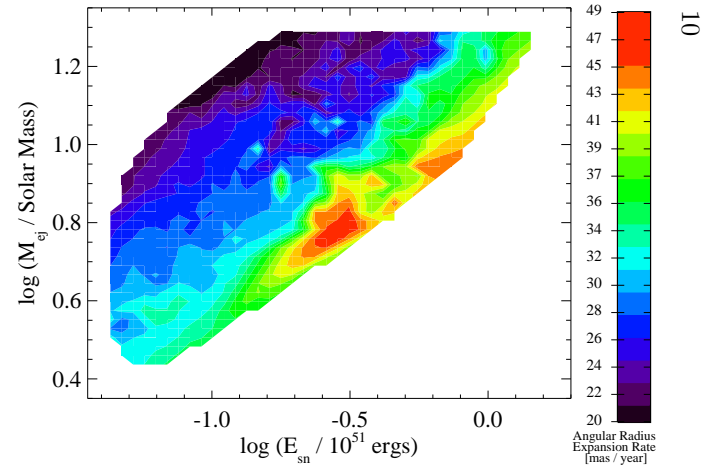} 
  \end{center}
  \caption{The expected angular expansion rate of this PWN's radius
    $\dot{\theta}_{\rm pwn}$ for different values of the initial
    kinetic energy $E_{\rm sn}$ and mass $M_{\rm ej}$ of the supernova
    ejecta for trials with $\chi^2<7.10$, $M_{\rm ej} < 20~M_{\odot}$,
    and $p<3$. }
  \label{fig:esn-mej_thetadotpwn}
\end{figure}

\begin{figure*}[tb]
  \begin{center}
    \includegraphics[width=0.49\textwidth]{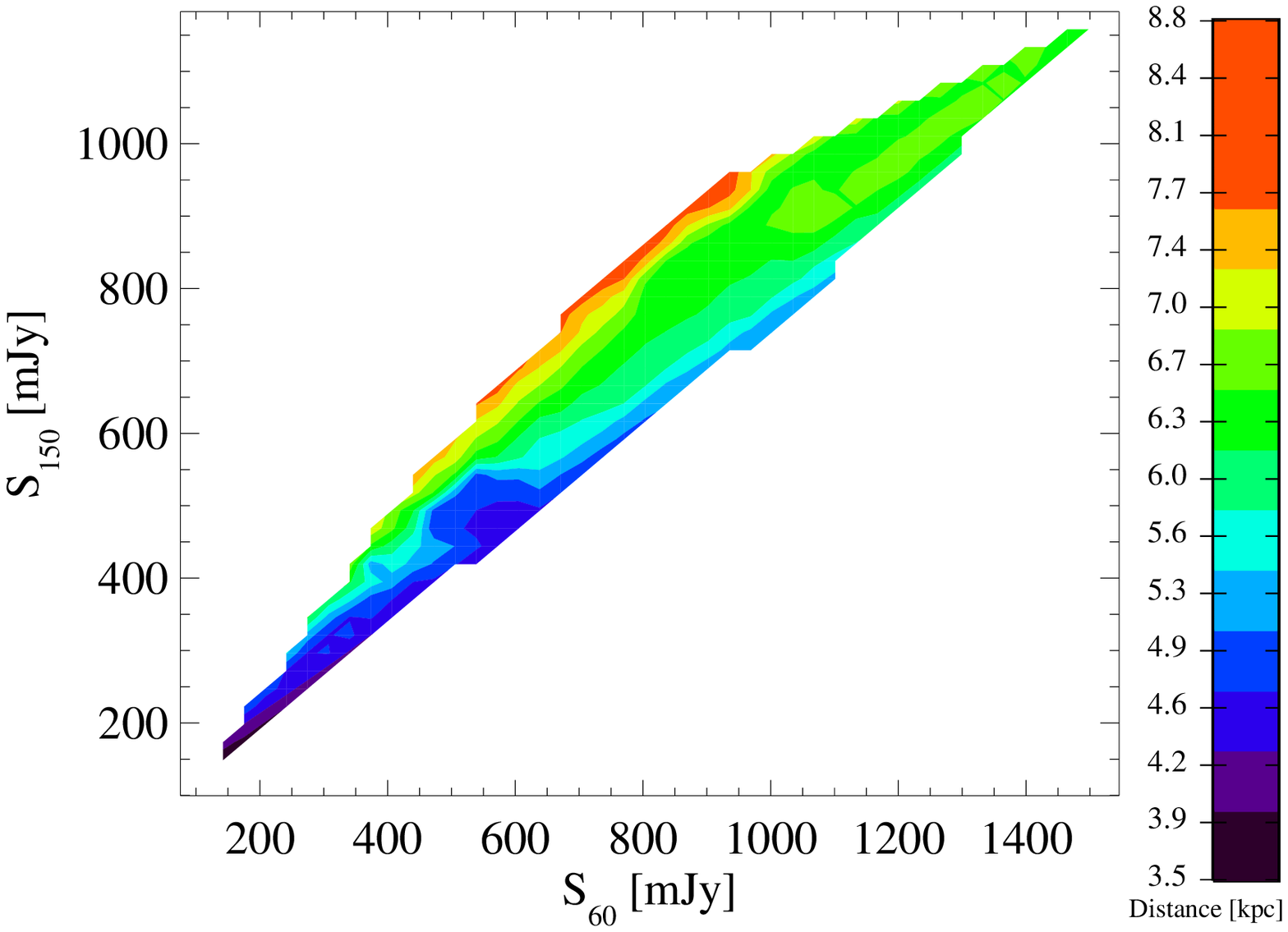}
    \includegraphics[width=0.49\textwidth]{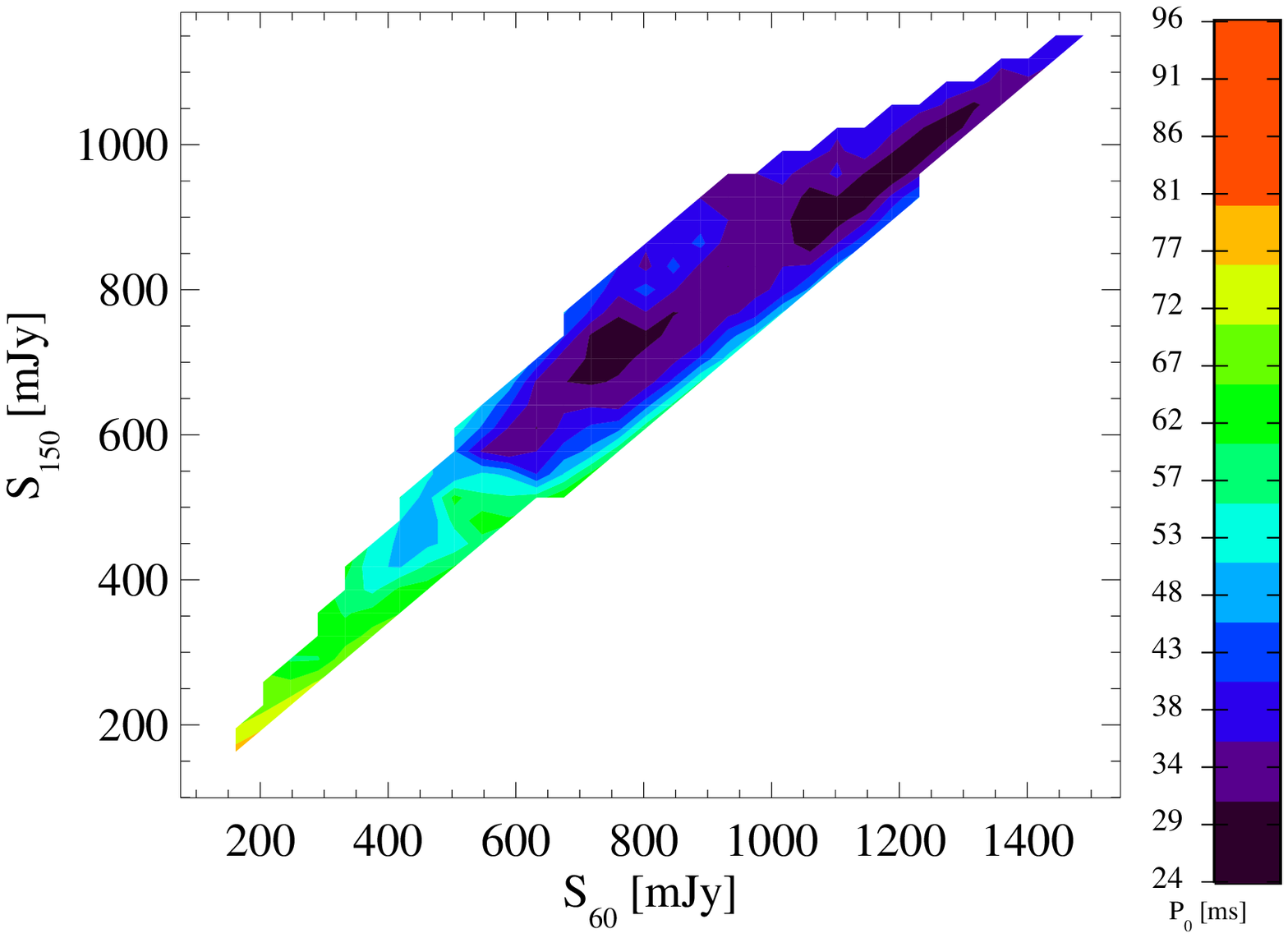}
  \end{center}
  \caption{The distance $d$ to G54.1+0.3 ({\it left}) and initial spin
    period $P_0$ of PSR J1930$+$1852 ({\it right}) for the values of
    the 60 MHz $S_{60}$ and 150 MHz $S_{150}$ flux densities predicted
    by trials with $\chi^2<7.10$, $M_{\rm ej} < 20~M_{\odot}$, and
    $p<3$.}
  \label{fig:p0-s60}
\end{figure*}

\begin{figure}[tb]
  \begin{center}
    \includegraphics[width=0.48\textwidth]{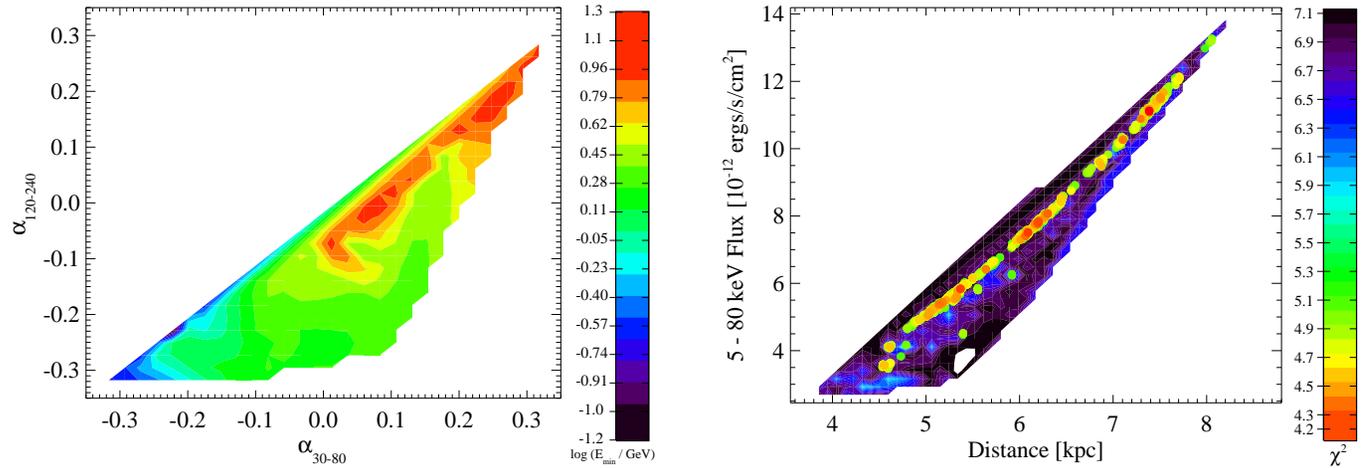}
  \end{center}
  \caption{The minimum energy in the pulsar wind $E_{\rm min}$ for the
    30$-$80 MHz $(\alpha_{30-80})$ and 120$-$240 MHz
    $(\alpha_{120-240})$ spectral indices of G54.1+0.3 predicted by
    trials with $\chi^2<7.10$, $M_{\rm ej} < 20~M_{\odot}$, and
    $p<3$.}
  \label{fig:alphalba}
\end{figure}

\begin{figure}[tb]
  \begin{center}
    \includegraphics[width=0.48\textwidth]{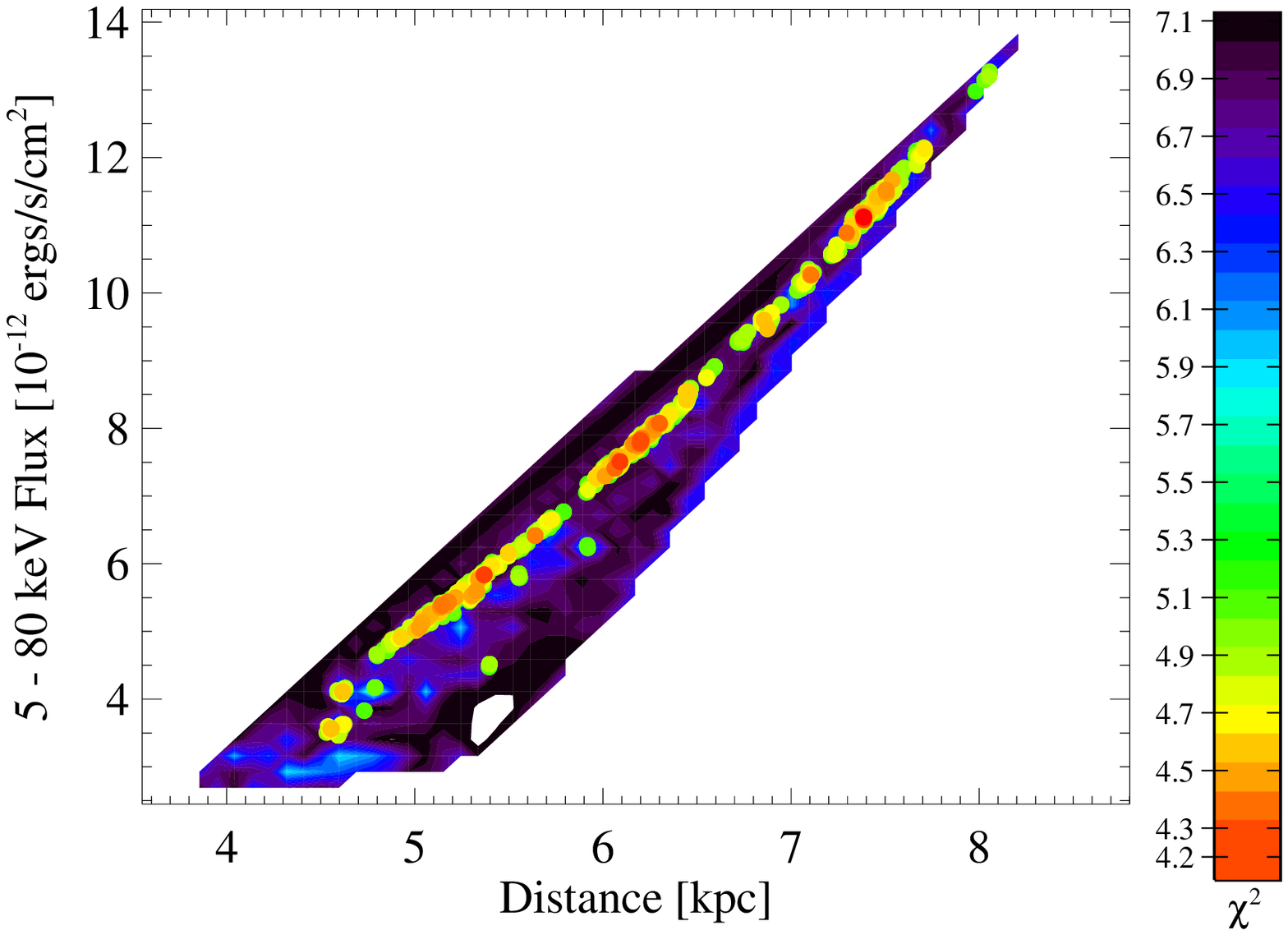}
  \end{center}
  \caption{The $\chi^2$ for trials with different distances $d$ and
    predicted 5$-$80 keV flux, which red signifying a lower $\chi^2$
    (better fit) and black a higher $\chi^2$ (worse fit).  Only the
    results of trials with $\chi^2<7.10$, $M_{\rm ej} < 20~M_{\odot}$,
    and $p<3$ are shown, and the dots indicate trials with
    $\chi^2 < 5.10$ and are included to better demonstrate the
    degeneracy between these two parameters.  The clumpiness of these
    points primarily reflects the sampling of the parameter space by
    our MCMC algorithm.}
\label{fig:hardxflux}
\end{figure}

We can also predict currently unmeasured properties of the PWN, and
determine what can be gained from their measurement.  For example, our
model predicts the average angular radius of the PWN is expanding by
$\dot{\theta}_{\rm pwn} \sim 20 - 50~{\rm mas~year}^{-1}$, and this
value is sensitive to the mass $M_{\rm ej}$ and initial kinetic energy
$E_{\rm sn}$ of the supernova ejecta because the PWN has not yet
collided with the SNR reverse shock (Figure
\ref{fig:esn-mej_thetadotpwn}).  This is potentially measurable using
high-resolution radio observations $\sim5-10$ years apart, though is
complicated by the considerable asymmetry of this PWN \citep{lang10}.
Additionally, we find that the flux density of G54.1+0.3 at low
frequencies (e.g., at 60~MHz $S_{60}$ and 150~MHz $S_{150}$) are
sensitive to both the distance $d$ to G54.1+0.3 and the initial spin
period $P_0$ of its associated pulsar PSR J1930+1852 (Figure
\ref{fig:p0-s60}).  Furthermore, the spectral indices in this band,
e.g. between $30-80$~MHz ($\alpha_{30-80}$) and $120-240$ MHz
($\alpha_{120-240}$) are sensitive to the minimum energy of particles
injected into the PWN at the termination shock $E_{\rm min}$ (Figure
\ref{fig:alphalba}).  All four of these quantities are measurable by
new observing facilities such as LOFAR \citep{lofar}.  Lastly, we find
that the absorbed $5-80$ keV flux of G54.1+0.3, measurable by the {\it
  NuSTAR} satellite \citep{nustar}, is strongly depends with the
distance to this source (Figure \ref{fig:hardxflux}) -- likely a
result of the parameter degeneracies discussed in Section
\ref{fitdesc}.

\section{Summary and Conclusions}
\label{conclusion}

In summary, we have fit the observed properties of G54.1+0.3 using a
one-zone model for the evolution of a PWN inside an SNR (Section
\ref{modeldesc}).  This model can reproduce its observed properties
(Section \ref{obsdata}), and suggests that the progenitor was an
isolated $\sim15~M_\odot$ star, most likely the member of a massive
star cluster, which exploded in a low density environment possibly
produced by its stellar wind (Section \ref{progenitor}).  The
resultant neutron star, PSR J1930$+$1852, had an initial spin period
$P_0 \sim 30-80~{\rm ms}$ (Section \ref{nsform}).  Our model requires
that the current multiplicity of particle production in its
magnetosphere is $\kappa \sim (1-3)\times10^5$, and suggests that the
magnetosphere electric potential is sufficient to accelerate particles
to the highest energies $E_{\rm max}$ required by our model.  The low
magnetization of the pulsar wind and low-energy component of particle
spectrum can be attributed to acceleration resulting from magnetic
reconnection between the light cylinder and the termination shock,
though our model suggests the ``stripes'' in the unshocked pulsar wind
are too narrow for acceleration at the termination shock to produce
the broken power-law spectrum required by our modeling.  These results
can be tested with radio and X-ray observations of this source, which
can better determine the initial spin period PSR J1930$+$1852, the
properties of particles accelerated in this source, and the nature of
the extended radio and X-ray emission surrounding this PWN.

\acknowledgements JDG will like to thank Erin Sheldon for the IDL code
used in the MCMC fits, Kaisey Mandel and David Hogg for useful
discussions concerning MCMC fitting, Ester Aliu for information
regarding the GeV spectrum, and Roger Chevalier, Vikram Dwarkadas,
Daniel Patnaude, and Lorenzo Sironi for useful advice.

\bibliography{ms}
\end{document}